\documentclass{iau307}
\usepackage{graphicx}
\usepackage{natbib}
\bibpunct{(}{)}{;}{a}{}{,}

\title[Combining seismology and spectropolarimetry] %% give here short title %%
{Combining seismology and spectropolarimetry of hot stars}

\author[C. Neiner et al.]   %% give here short author list %%
{Coralie Neiner$^1$,
Maryline Briquet$^{2,1}$,
St\'ephane Mathis$^{3,1}$
\and
Pieter Degroote$^{4,1}$
}

\affiliation{$^1$LESIA, Observatoire de Paris, CNRS UMR 8109, UPMC, Universit\'e
Paris Diderot,\\
5 place Jules Janssen, 92190 Meudon, France \\
email: {\tt coralie.neiner@obspm.fr}\\[\affilskip]
$^2$ Institut d'Astrophysique et de G\'eophysique, Universit\'e de Li\`ege,\\
All\'ee du 6 Ao\^ut 17, B\^at B5c, 4000 Li\`ege, Belgium
\\[\affilskip]
$^3$ Laboratoire AIM Paris-Saclay, CEA/DSM - CNRS - Universit\'e Paris
Diderot,\\
IRFU/SAp Centre de Saclay, 91191 Gif-sur-Yvette, France
\\[\affilskip]
$^4$ Instituut voor Sterrenkunde, Celestijnenlaan 200D, 3001 Heverlee, 
Belgium
}

\pubyear{2014}
\volume{307} 
\pagerange{}
\jname{New windows on massive stars: asteroseismology, interferometry, and spectropolarimetry}
\editors{G. Meynet, C. Georgy, J.H. Groh \& Ph. Stee, eds.}
\begin{document}

\maketitle

\begin{abstract}
Asteroseismology and spectropolarimetry have allowed us to progress
significantly in our understanding of the physics of hot stars over the last
decade. It is now possible to combine these two techniques to learn even more
information about hot stars and constrain their models. While only a few
magnetic pulsating hot stars are known as of today and have been studied with
both seismology and spectropolarimetry, new opportunities - in particular
{\it Kepler}2 and BRITE - are emerging and will allow us to rapidly obtain new
combined results.
\end{abstract}

\firstsection % if your document starts with a section,
\section{Introduction}

Over the last decade, two major steps have been done in parallel leading to
substantial progress in the field of hot stars:\\
(1) asteroseismology, in particular with space-based facilities such as MOST, 
CoRoT and {\it Kepler} but also with multi-site spectroscopic campaigns, has
allowed us to study the pulsations of hot stars with high precision and thus
their internal structure (see Aerts, this volume),\\
(2) spectropolarimetry, with the new generation of high-resolution instruments
Narval at TBL, ESPaDOnS at CFHT and HarpsPol at ESO, has allowed us to study the
magnetic fields of hot stars and their circumstellar magnetospheres (see
Grunhut, this volume).\\
We have now reached a point where these two techniques can be and should be
combined to increase the physics we can probe in hot stars.

About 10\% of hot stars are found to be magnetic with oblique dipolar fields
above $\sim$100 G at the pole \citep{wade2013}. A similar occurence of magnetic
fields is observed in pulsating hot stars and in non-pulsating ones
\citep{neiner2011}. The presence of this magnetic field impacts the power
spectrum of pulsations that we observe in $\beta$\,Cep, Slowly Pulsating B (SPB)
and roAp stars. In particular magnetic splitting of the pulsation modes occurs.
The width of this splitting is directly related to the strength of the field,
but the amplitude of the components of the splitted multiplet depends on the
obliquity of the dipolar magnetic field compared to the rotation/pulsation axis
\citep[see e.g.][]{shibahashi2000}. Therefore the number of peaks that can be
detected in the power spectrum with current instrumentation depends on the
magnetic geometry. Knowing this information is very important to perform a
proper pulsation mode identification and thus compute a correct seismic model.

Conversely, knowing the pulsation properties of a hot star is important when
studying its magnetic field. Indeed, the Zeeman Stokes signature of the magnetic
field observed in spectropolarimetry, which allows us to derive the longitudinal
field value and magnetic configuration, depends on the intensity of the line
profiles and thus on how the lines are deformed by pulsations. In particular,
variations of the line profiles due to pulsations should be taken into account
when measuring the longitudinal magnetic field. To derive proper magnetic field
strength and geometry, one thus needs to model the intensity line profiles
taking pulsations into account.

\section{Magnetic modelling of pulsating stars}

The Phoebe2.0 code \citep{degroote2013} allows users to model various surface
observables  such as the light curve or line profiles of a star with pulsations,
spots, one or more companions, etc. The code has recently been modified to also
provide the possibility to include an oblique magnetic dipole field at the
surface of the modeled star.

Thanks to this code, we have modeled the line profile variations due to
pulsations of the star $\beta$\,Cep and the corresponding Stokes V profiles due
to the magnetic field \citep[see][]{neiner2013}.

$\beta$\,Cep is the first pulsating B star discovered to host a magnetic field
\citep{henrichs2000} and its field configuration has thus been well studied
since then. \cite{donati2001} proposed that B$_{\rm pol}$=360 G, the inclination
angle i is 60$^\circ$ and the obliquity angle $\beta$ is 85$^\circ$.
\cite{henrichs2013} showed that B$_{\rm pol}>$306 G (since B$_{\rm max}$=97 G),
with the same inclination angle i=60$^\circ$ but $\beta$=96$^\circ$. 

We have used 56 new Narval observations of $\beta$\,Cep and applied the LSD
technique \citep{donati1997} to produce LSD Stokes I and V profiles. The I
profiles vary mostly with the pulsation periods, while the Stokes V profiles
clearly vary with both the pulsation periods and rotation period (because of
the rotational modulation due to the field obliquity).

We fitted these LSD I and V profiles with Phoebe2.0. Our model results in
B$_{\rm pol}$=370 G and $\beta$=92$^\circ$, compatible with the values published
in the literature when ignoring pulsations. However, the most striking result is
that the inclination angle we find is different from the one published so far.
We find that i=46$^\circ$ rather than i=60$^\circ$ (Neiner et al., in prep.).

This shows that ignoring pulsations in magnetic studies of pulsating stars
introduces large errors in the geometrical configuration that is deduced from
the Stokes profiles. Therefore, it is necessary to take pulsations into account,
especially when their amplitude is large and in the case of radial modes, to
obtain the proper geometry. 

\section{Seismic modelling of magnetic stars}

The second pulsating B star discovered to host a magnetic field is V2052\,Oph
\citep{neiner2003}. Its magnetic field was recently analysed in more details
thanks to new Narval observations \citep{neiner2012}. It was found that the
magnetic field of V2052\,Oph has a polar strength of B$_{\rm pol}\sim$400 G with
an obliquity of $\beta\sim35^\circ$ and an inclination of i$\sim70^\circ$.

A multi-site spectroscopic \citep{briquet2012} and photometric
\citep{handler2012} campaign also allowed us to reanalyse its pulsations in more
details. The stellar pulsations are dominated by a radial mode but two
non-radial low-amplitude prograde modes are also detected. Rotational modulation
is also clearly present due to the oblique magnetic field. Moreover, we found
that the seismic models that reproduce best the observations are those in which
no or very small overshooting is added ($\alpha$ = 0.07 $\pm$ 0.08 H$_{\rm p}$).
Such a low overshooting parameter is not common in $\beta$\,Cep stars especially
when they rotate as fast as V2052\,Oph ($v\sin{i}$=80 km~s$^{-1}$) and when
rotational mixing is thus expected.

The reason for this lack of macroscopic internal mixing is the presence of the
magnetic field. Indeed, using the criteria proposed by \cite{zahn2011} based on
\cite{mathis2005} or by \cite{spruit1999}, we derive that mixing is inhibited
in V2052\,Oph when B$_{\rm pol}\sim$70 G and 40 G, respectively
\citep{briquet2012}. This critical field value is 6 to 10 times weaker than the
observed polar field strength of V2052\,Oph. Therefore, the seismic model of
V2052\,Oph is the first observational proof that internal mixing is inhibited by
magnetic field.

The criteria proposed by \cite{zahn2011} and \cite{spruit1999}, however, use
several crude approximations. We have thus started to devise a new more precise
critical field criterion, computing the magnetic torque with realistic
non-axisymmetric geometry, i.e. taking obliquity into account (Mathis et al., in
prep.). This criterion is of the form:

\begin{equation}
B_{\rm crit}^2 = 4 \pi \rho R^2 \Omega / t_{\rm AM} * F(\beta),
\end{equation}
where $\rho$ is the density of the star, R is its radius, $\Omega$ is its
angular velocity, $t_{\rm AM}$ is the characteristic time for the angular
momentum evolution (due to the wind or to structural adjustments), and $F$ is a
function of the obliquity angle  $\beta$.

We will apply this criterion to all known magnetic hot stars. This will allow us
to provide a new constraint for seismic models from spectropolarimetry, i.e. to
tell if overshoot should be added in the seismic models or fixed to 0.
Spectropolarimetry also already allows us to constraint these models by
providing the geometry and in particular the inclination angle.

\section{New opportunities for seismic studies of magnetic stars}

\begin{figure}[t]
\begin{center}
\includegraphics[width=\textwidth]{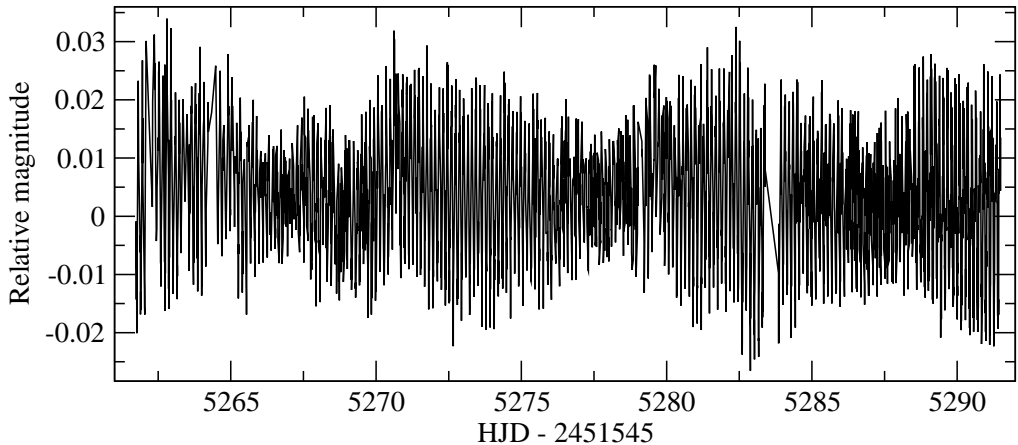} 
\caption{MOST light curve of the magnetic pulsating B star V2052\,Oph.}
\label{most_v2052oph}
\end{center}
\end{figure}

MOST observed several bright pulsating OB stars and a few magnetic hot stars,
but had not observed any pulsating magnetic hot star so far. While MOST will
ceased its observations at the end of August 2014, one of its last target from
May 29 to June 27, 2014, was V2052\,Oph. Therefore a MOST light curve of this
well-studied magnetic pulsating star has just become available (see
Fig.~\ref{most_v2052oph}). We hope to be able to identify magnetic splittings in
the power spectrum of the pulsations of this star.

CoRoT observed many OB stars. None of them were known to be magnetic at the time
of observations. However, targets observed in the ``seismo'' fields of CoRoT are
sufficiently bright to be observed with high-resolution spectropolarimetry. This
was done with Narval for all hot stars. Among these targets, one of them,
HD\,43317, was found to be magnetic \citep{briquet2013}. HD\,43317 is a hybrid
($\beta$\,Cep + SPB) pulsator and thus exhibits both p and g pulsation modes. In
addition, regular frequency spacing was detected in its power spectrum
\citep{papics2012}. A first seismic model was computed \citep{savonije2013}.
However, the subsequent discovery of the magnetic field in HD\,43317 calls for a
reinterpretation of the observations, in particular in terms of magnetic
splittings, and a new seismic model taking the magnetic field into account. The
fact that HD\,43317 shows p modes, g modes, rotational modulation and a magnetic
field provides a wealth of constraints for such a model.

Finally, {\it Kepler} observed only one field of view optimised for the
detection of exoplanets around solar-type stars. This field unfortunately did
not contain O stars nor early-B stars. It did contain some late-B stars and thus
SPB stars, but none of them are known to be magnetic and these targets are too
faint for current spectropolarimeters to detect a field unless it would be many
kG in strength.

As a consequence, only very few magnetic pulsating stars could be observed so
far with space-based photometry. New opportunities are however opening now: {\it
Kepler}2 and BRITE.

\subsection{{\it Kepler}2}

Following the failure of several reaction wheels in {\it Kepler}, the
spacecraft's objectives have been redefined into a new mission called {\it
Kepler}2 (or K2). K2 is limited to pointing fields near the ecliptic plane and
has to change field about every 3 months. As a consequence, the new fields of
K2 include all kinds of targets.

In particular, Field 0 included 35 magnetic hot stars, among which 7 were
observed between March 8 and May 30, 2014. Field 2 also contains 5 magnetic hot
stars, which have been requested for observation in the second semester of 2014.
Several magnetic hot stars will also be observable in the forthcoming Fields 4,
5, 7 and 9.

These K2 data will multiply tenfold the number of magnetic hot stars observed 
with high-precision space photometry. Many of these targets are expected to
pulsate and will thus provide a unique opportunity for combined seismic and
spectropolarimetric studies. 

\subsection{BRITE}

The BRIght Target Explorer (BRITE) is an Austrian, Polish and Canadian
constellation of nano-satellites \citep{weiss2014}. Each participating country
builds two nano-satellites: one providing photometry in the red band and the
other one in the blue band. Using two colours allows one to more easily identify
the pulsation modes of hot stars thanks to the difference in mode amplitude in
the two colours.

The constellation is already composed of two Austrian, one Polish and one
Canadian nano-satellites. The second Polish nano-satellite should be launched
soon, while the second Canadian nano-satellite unfortunately failed to separate
from the upper-stage of the launch vehicle during its launch in June 2014.

BRITE concentrates on targets brighter than V=4. As a consequence it mostly
observes hot stars and evolved stars. There are $\sim$600 stars with V$\le$4,
among which $\sim$300 are hotter than F5, i.e. are stars in which a fossil
magnetic field can be detected.

As of now, we know 8 magnetic pulsating OB stars with V$\le$4:
$\beta$\,Cen, $\zeta$\,Ori\,A, $\tau$\,Sco, $\beta$\,Cep, $\epsilon$\,Lup,
$\zeta$\,Cas, $\phi$\,Cen and $\beta$\,CMa.

Since BRITE targets are very bright, they can very easily be observed in
spectropolarimetry with a very good magnetic field detection threshold. We have
thus started a large survey of all stars with V$\le$4 with Narval (for stars
with a declination $\delta>-20^\circ$; PI Neiner), ESPaDOnS (for stars with
$-45<\delta<-20^\circ$; PI Wade) and HarpsPol (for stars with
$\delta<-45^\circ$; PI Neiner). We aim at a detection threshold of B$_{\rm
pol}$=50 G for all fossil field stars i.e. hotter than F5, and B$_{\rm pol}$=5 G
for cooler stars with a magnetic field of dynamo origin. These very high quality
data will also provide an ideal spectroscopic database, e.g. for the
determination of the stellar parameters of all BRITE targets.

The spectropolarimetric observations of BRITE targets started in February 2014.
We have already discovered 9 new magnetic stars. However, these are mainly cool
stars. The hottest one, $\iota$\,Peg, is a F5V star and a SB2 spectroscopic
binary.

\section{Conclusion}

Combined magnetic and seismic observational studies can now be performed on
bright hot stars. Such studies provide better and stronger constraints for
models, e.g. on internal mixing or on the geometrical configuration of the star
(inclination and field obliquity). They thus allow us to better understand the
physics at work inside hot stars. 

Conversely, ignoring one or the other aspect (pulsations or magnetism) can lead
to substantially wrong results, therefore combined studies should be performed
whenever possible. To this aim, {\it Kepler}2 and BRITE will provide new
opportunities. 

\bibliographystyle{iau307}
\bibliography{IAUS307Neiner1}

\begin{thebibliography}{}

\bibitem[\protect\astroncite{{Briquet} et~al.}{2012}]{briquet2012}
{Briquet}, M., {Neiner}, C., {Aerts}, C., {et~al.} 2012,
\newblock {\em \mnras} 427, 483

\bibitem[\protect\astroncite{{Briquet} et~al.}{2013}]{briquet2013}
{Briquet}, M., {Neiner}, C., {Leroy}, B., \& {P{\'a}pics}, P.~I. 2013,
\newblock {\em \aap} 557, L16

\bibitem[\protect\astroncite{{Degroote} et~al.}{2013}]{degroote2013}
{Degroote}, P., {Conroy}, K., {Hambleton}, K., {et~al.} 2013,
\newblock in {\em EAS Publications Series}, Vol.~64 of {\em EAS Publications
  Series}, p. 277

\bibitem[\protect\astroncite{{Donati} et~al.}{1997}]{donati1997}
{Donati}, J.-F., {Semel}, M., {Carter}, B.~D., {Rees}, D.~E., \& {Collier
  Cameron}, A. 1997,
\newblock {\em \mnras} 291, 658

\bibitem[\protect\astroncite{{Donati} et~al.}{2001}]{donati2001}
{Donati}, J.-F., {Wade}, G.~A., {Babel}, J., {et~al.} 2001,
\newblock {\em \mnras} 326, 1265

\bibitem[\protect\astroncite{{Handler} et~al.}{2012}]{handler2012}
{Handler}, G., {Shobbrook}, R.~R., {Uytterhoeven}, K., {et~al.} 2012,
\newblock {\em \mnras} 424, 2380

\bibitem[\protect\astroncite{{Henrichs} et~al.}{2000}]{henrichs2000}
{Henrichs}, H.~F., {de Jong}, J.~A., {Donati}, J.-F., {et~al.} 2000,
\newblock in M.~A. {Smith}, H.~F. {Henrichs}, \& J. {Fabregat} (eds.), {\em IAU
  Colloq. 175: The Be Phenomenon in Early-Type Stars}, Vol. 214 of {\em
  Astronomical Society of the Pacific Conference Series}, p. 324

\bibitem[\protect\astroncite{{Henrichs} et~al.}{2013}]{henrichs2013}
{Henrichs}, H.~F., {de Jong}, J.~A., {Verdugo}, E., {et~al.} 2013,
\newblock {\em \aap} 555, A46

\bibitem[\protect\astroncite{{Mathis} \& {Zahn}}{2005}]{mathis2005}
{Mathis}, S. \& {Zahn}, J.-P. 2005,
\newblock {\em \aap} 440, 653

\bibitem[\protect\astroncite{{Neiner} et~al.}{2012}]{neiner2012}
{Neiner}, C., {Alecian}, E., {Briquet}, M., {et~al.} 2012,
\newblock {\em \aap} 537, A148

\bibitem[\protect\astroncite{{Neiner} et~al.}{2011}]{neiner2011}
{Neiner}, C., {Alecian}, E., \& {Mathis}, S. 2011,
\newblock in G. {Alecian}, K. {Belkacem}, R. {Samadi}, \& D. {Valls-Gabaud}
  (eds.), {\em SF2A-2011: Proceedings of the Annual meeting of the French
  Society of Astronomy and Astrophysics}, p. 509

\bibitem[\protect\astroncite{{Neiner} et~al.}{2013}]{neiner2013}
{Neiner}, C., {Degroote}, P., {Coste}, B., {Briquet}, M., \& {Mathis}, S. 2013,
\newblock in {\em Magnetic fields throughout stellar evolution}, Vol. 302 of
  {\em IAU Symposium, ArXiv 1311.2262}

\bibitem[\protect\astroncite{{Neiner} et~al.}{2003}]{neiner2003}
{Neiner}, C., {Henrichs}, H.~F., {Floquet}, M., {et~al.} 2003,
\newblock {\em \aap} 411, 565

\bibitem[\protect\astroncite{{P{\'a}pics} et~al.}{2012}]{papics2012}
{P{\'a}pics}, P.~I., {Briquet}, M., {Baglin}, A., {et~al.} 2012,
\newblock {\em \aap} 542, A55

\bibitem[\protect\astroncite{{Savonije}}{2013}]{savonije2013}
{Savonije}, G.~J. 2013,
\newblock {\em \aap} 559, A25

\bibitem[\protect\astroncite{{Shibahashi} \& {Aerts}}{2000}]{shibahashi2000}
{Shibahashi}, H. \& {Aerts}, C. 2000,
\newblock {\em \apjl} 531, L143

\bibitem[\protect\astroncite{{Spruit}}{1999}]{spruit1999}
{Spruit}, H.~C. 1999,
\newblock {\em \aap} 349, 189

\bibitem[\protect\astroncite{{Wade} et~al.}{2013}]{wade2013}
{Wade}, G.~A., {Grunhut}, J., {Alecian}, E., {et~al.} 2013,
\newblock in {\em Magnetic fields throughout stellar evolution}, Vol. 302 of
  {\em IAU Symposium, ArXiv 1310.3965}

\bibitem[\protect\astroncite{{Weiss} et~al.}{2014}]{weiss2014}
{Weiss}, W.~W., {Rucinski}, S.~M., {Moffat}, A.~F.~J., {et~al.} 2014,
\newblock {\em \pasp} 126, 573

\bibitem[\protect\astroncite{{Zahn}}{2011}]{zahn2011}
{Zahn}, J.-P. 2011,
\newblock in C. {Neiner}, G. {Wade}, G. {Meynet}, \& G. {Peters} (eds.), {\em
  IAU Symposium}, Vol. 272 of {\em IAU Symposium}, p.~14

\end{thebibliography}

\begin{discussion}

\discuss{Huib Henrichs}{Congratulations with this beautiful work. Now that
there is a new determination of the inclination angle of $\beta$\,Cep, I wonder
if the UV wind behavior with its assymmetry could be modelled as well. In
particular the assymetry of the equivalent width at opposite orientation
pointed to an off-centered wind configuration. Would this be reproduced? And
what about the wind profiles themselves?}

\discuss{Coralie Neiner}{In the Phoebe2.0 model of $\beta$\,Cep, we introduced
as many observational constraints as possible. This includes not only the
variable intensity line profiles and the magnetic Stokes V profiles, but also,
e.g., constraints on the stellar parameters such as the distance or mass
obtained thanks to speckle interferometry and radial velocities. As far as the
UV is concerned, we have used the rotational modulation that you derived from
the UV wind variations, but we have not modeled the wind line profiles
themselves. Our model points to a mainly dipolar (but oblique) magnetic field.}

\discuss{Andr\'e Maeder}{What may be the typical value of this new factor $F$
that accounts for the obliquity in the expression of the critical field?}

\discuss{Coralie Neiner}{This factor is composed of trigonometric functions of
the obliquity $\beta$ corresponding to the projection of the Lorentz force in
the reference frame associated to the rotation axis. The exact details of this
factor are still under investigation by St\'ephane Mathis. } 

\end{discussion}

\end{document}